\documentclass{article}

\usepackage{microtype}
\usepackage{graphicx}
\usepackage{subfigure}
\usepackage{booktabs} % for professional tables

\usepackage{xcolor} % For color definitions
\usepackage{colortbl} % For coloring table rows

\usepackage{hyperref}

\usepackage[accepted]{lxai}

\usepackage{amsmath}
\usepackage{amssymb}
\usepackage{mathtools}
\usepackage{amsthm}

\usepackage[capitalize,noabbrev]{cleveref}

\usepackage{tabularx}
\usepackage{ragged2e} % For better text alignment in tabularx
\theoremstyle{plain}

\theoremstyle{definition}

\theoremstyle{remark}

\usepackage[textsize=tiny]{todonotes}

\icmltitlerunning{Submission and Formatting Instructions for ICML 2025}

\begin{document}

\twocolumn[
%\icmltitle{Decolonizing AI: Evaluating Cultural Expressiveness in LLMs for Latin America}
\icmltitle{Advancing Equitable AI: Evaluating Cultural Expressiveness in LLMs for Latin American Contexts}

% It is OKAY to include author information, even for blind
% submissions: the style file will automatically remove it for you
% unless you've provided the [accepted] option to the icml2025
% package.

% List of affiliations: The first argument should be a (short)
% identifier you will use later to specify author affiliations
% Academic affiliations should list Department, University, City, Region, Country
% Industry affiliations should list Company, City, Region, Country

% You can specify symbols, otherwise they are numbered in order.
% Ideally, you should not use this facility. Affiliations will be numbered
% in order of appearance and this is the preferred way.
\icmlsetsymbol{equal}{*}

\begin{icmlauthorlist}
\icmlauthor{Brigitte A. Mora-Reyes}{equal,ug}
\icmlauthor{Jennifer A. Drewyor}{equal,mtu_ps}
\icmlauthor{Abel A. Reyes-Angulo}{mtu_ac}
%\icmlauthor{Firstname4 Lastname4}{sch}
%\icmlauthor{Firstname5 Lastname5}{yyy}
%\icmlauthor{Firstname6 Lastname6}{sch,yyy,comp}
%\icmlauthor{Firstname7 Lastname7}{comp}
%\icmlauthor{}{sch}
%\icmlauthor{Firstname8 Lastname8}{sch}
%\icmlauthor{Firstname8 Lastname8}{yyy,comp}
%\icmlauthor{}{sch}
%\icmlauthor{}{sch}
\end{icmlauthorlist}

\icmlaffiliation{ug}{Facultad Jurisprudencia Ciencias Sociales y Políticas, Universidad de Guayaquil, Guayaquil, Ecuador}
\icmlaffiliation{mtu_ac}{Department of Applied Computing, Michigan Technological University, Houghton, MI, USA}
\icmlaffiliation{mtu_ps}{Department of Psychology and Human Factors, Michigan Technological University, Houghton, MI, USA}

\icmlcorrespondingauthor{Brigitte A. Mora-Reyes}{brigitte.morar@ug.edu.ec}
\icmlcorrespondingauthor{Jennifer A. Drewyor}{jadrewyo@mtu.edu}

% You may provide any keywords that you
% find helpful for describing your paper; these are used to populate
% the "keywords" metadata in the PDF but will not be shown in the document
\icmlkeywords{Machine Learning, ICML}

\vskip 0.3in
]

% this must go after the closing bracket ] following \twocolumn[ ...

% This command actually creates the footnote in the first column
% listing the affiliations and the copyright notice.
% The command takes one argument, which is text to display at the start of the footnote.
% The \icmlEqualContribution command is standard text for equal contribution.
% Remove it (just {}) if you do not need this facility.

%\printAffiliationsAndNotice{}  % leave blank if no need to mention equal contribution
\printAffiliationsAndNotice{\icmlEqualContribution} % otherwise use the standard text.

% \begin{abstract}
% Artificial intelligence (AI) systems often reflect Global North biases, marginalizing Latin American contexts due to imbalanced datasets. This paper examines AI representations of Latin American issues, revealing disparities between Global North and South data. We highlight how the dominance of English over Spanish, Portuguese, and indigenous languages like Quechua perpetuates biases, framing Latin American issues through a Western lens. To address this, we introduce a culturally aware dataset grounded in Latin American history and socio-political contexts, challenging Eurocentric models. We evaluate six language models on questions testing cultural context awareness, using a novel Cultural Expressiveness metric, statistical tests, and linguistic analyses. Our findings show that some models better capture Latin American perspectives, while others exhibit significant sentiment misalignment (\( p < 0.001 \)). Fine-tuning Mistral-7B with our dataset improves its cultural expressiveness by 42.9\%, a step toward equitable AI. We advocate for decolonizing AI by prioritizing datasets reflecting Latin American history and indigenous knowledge, offering a scalable framework to amplify marginalized voices. Code and datasets will be available in a public repository.
% \end{abstract}

\begin{abstract}
Artificial intelligence (AI) systems often reflect biases from economically advanced regions, marginalizing contexts in economically developing regions like Latin America due to imbalanced datasets. This paper examines AI representations of diverse Latin American contexts, revealing disparities between data from economically advanced and developing regions. We highlight how the dominance of English over Spanish, Portuguese, and indigenous languages such as Quechua and Nahuatl perpetuates biases, framing Latin American perspectives through a Western lens. To address this, we introduce a culturally aware dataset rooted in Latin American history and socio-political contexts, challenging Eurocentric models. We evaluate six language models on questions testing cultural context awareness, using a novel Cultural Expressiveness metric, statistical tests, and linguistic analyses. Our findings show that some models better capture Latin American perspectives, while others exhibit significant sentiment misalignment (\(p < 0.001\)). Fine-tuning Mistral-7B with our dataset improves its cultural expressiveness by 42.9\%, advancing equitable AI development. We advocate for equitable AI by prioritizing datasets that reflect Latin American history, indigenous knowledge, and diverse languages, while emphasizing community-centered approaches to amplify marginalized voices. Code and datasets available at: \href{https://github.com/areyesan/Advancing-Equitable-AI}{https://github.com/areyesan/Advancing-Equitable-AI}%will be available in a public repository.
\end{abstract}

\section{Introduction}
\label{introduction}

Artificial intelligence (AI) systems, particularly large language models (LLMs), often reflect biases rooted in datasets sourced from economically advanced regions, marginalizing economically developing regions like Latin America. This leads to incomplete or Eurocentric representations of diverse local contexts, such as indigenous rights, socio-political dynamics, and cultural identities across urban, rural, and indigenous communities, perpetuating cultural erasure (the loss or suppression of cultural identities due to dominant influences) and a center-periphery dynamic (a socio-economic framework where a dominant center exploits or marginalizes peripheral regions). Here, "economically advanced regions" refers to industrialized nations with high GDP (e.g., USA, Western Europe), while "economically developing regions" include Latin America and other areas with emerging economies. 

In this paper, we address these biases by introducing a culturally aware dataset tailored to diverse Latin American contexts, evaluating six LLMs—Mistral-7B, Zephyr-7B, BLOOM-7B, Llama-2-7B, Grok, and ChatGPT—to quantify their cultural and tonal disparities. Our analysis reveals that LLMs often fail to capture the linguistic diversity (e.g., Spanish, Portuguese, Quechua, Nahuatl) and socio-political nuances of Latin America, with some models exhibiting excessive positivity or negativity that misaligns with local perspectives. To bridge this gap, as shown in Figure 1, we propose a framework to enhance cultural expressiveness, detailed in Sections~\ref{sec:finetuning} and~\ref{sec:performance_improvement}, achieving a 42.9\% improvement (Section~\ref{sec:performance_improvement}). Our work contributes to advancing equitable AI by advocating for inclusive datasets that prioritize regional histories, indigenous knowledge, and local authorship, while emphasizing community-centered approaches for future development.

\begin{figure*}[t]
\centering
\includegraphics[width=1.0\textwidth]{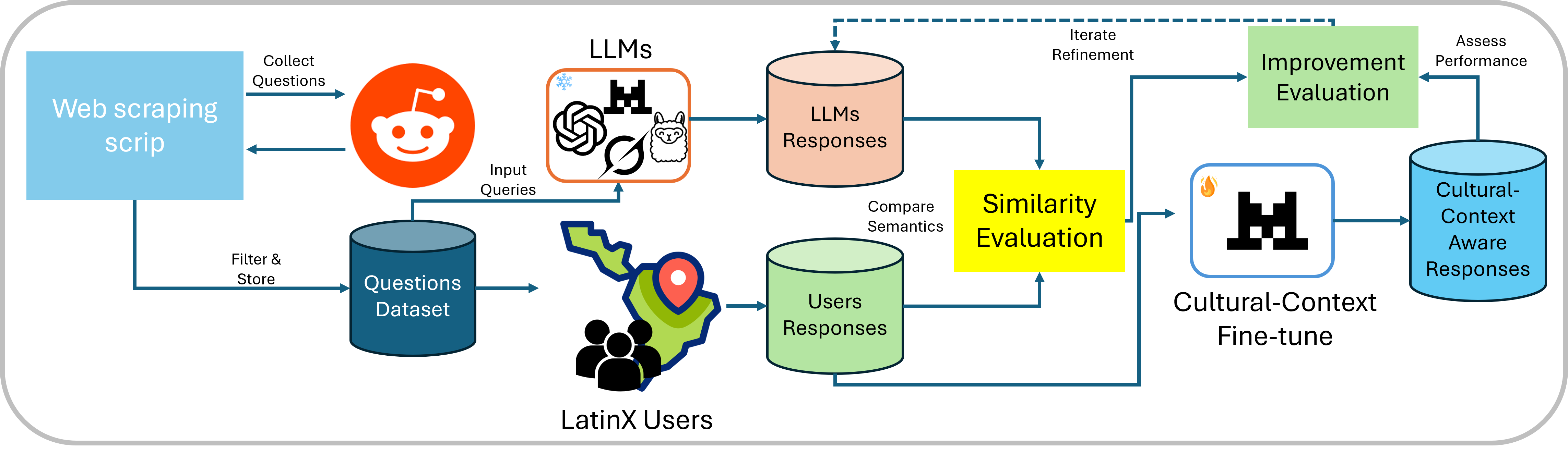}
\caption{Framework proposed to inject cultural context awareness into the knowledge of the LLMs.}
\label{fig:framework}
\end{figure*}

\section{Background}
\label{background}

The development and deployment of large language models (LLMs) have been transformative, yet their reliance on datasets skewed toward the Global North has raised concerns about cultural biases and representational harms. Bender \cite{bender2021dangers} argue that the homogeneity of training data leads to models that perpetuate stereotypes and marginalize underrepresented groups, including those from Latin America. Similarly, Blodgett \cite{blodgett2020language} emphasize that language technologies often fail to account for linguistic diversity, such as the dialects and indigenous languages prevalent in Latin America, resulting in outputs that lack cultural relevance.

Efforts to make AI equitable have gained traction, with researchers advocating for datasets and models that reflect diverse cultural realities. Sambasivan \cite{sambasivan2021re} proposes reimagining AI development through participatory approaches that center marginalized communities, such as those in the Global South. In the Latin American context, Ricaurte \cite{ricaurte2019data} highlights how AI perpetuates colonial biases, often misrepresenting regional issues like indigenous land rights due to imbalanced datasets. These works underscore the need for region-specific datasets that capture local knowledge and perspectives, a gap our culturally aware dataset aims to address.

Fine-tuning LLMs to incorporate cultural context has emerged as a promising strategy to mitigate biases. Techniques such as supervised fine-tuning (SFT) and reinforcement learning with human feedback (RLHF) have been used to align models with specific user needs or cultural expectations, as demonstrated by Ouyang et al. \cite{ouyang2022training} in their work on InstructGPT. Recent work by Hershcovich et al. \cite{hershcovich2022challenges} highlights challenges in cross-lingual and cross-cultural NLP, showing that targeted approaches can enhance a model’s ability to handle culturally nuanced contexts. However, few studies focus on Latin American contexts, where linguistic diversity and socio-political complexities demand tailored approaches.

Sentiment analysis and semantic similarity metrics have been widely used to evaluate LLM performance in cultural settings. Sanh et al. \cite{sanh2019distilbert} introduce DistilBERT, a lightweight model for sentiment analysis, which we leverage to assess tonal alignment between LLM and human responses. Semantic similarity, often measured via embeddings like those from Sentence-BERT \cite{reimers2019sentence}, provides a robust method to compare the contextual alignment of responses, as applied in our study to quantify cultural expressiveness.

\section{Methodology}
\label{methodology}
\subsection{Data Collection}
To build a culturally aware dataset reflective of Latin American perspectives, we collected questions from online forums using web scraping, with Reddit as the primary source due to its vibrant, region-specific communities. We targeted 13 subreddits focused on Latin American culture, history, and socio-political issues, including \texttt{r/AskLatinAmerica}, \texttt{r/LatinAmerica}, \texttt{r/Mexico}, \texttt{r/Brazil}, \texttt{r/Ecuador}, \texttt{r/Colombia}, \texttt{r/Peru}, \texttt{r/Venezuela}, \texttt{r/Chile}, \texttt{r/Argentina}, \texttt{r/Uruguay}, \texttt{r/Bolivia}, and \texttt{r/Paraguay}. These subreddits were chosen for their rich discussions, which capture the linguistic and cultural diversity of the region, spanning English, Spanish, and Portuguese, as well as references to indigenous languages like Quechua and Nahuatl in posts related to cultural identity and history.

The web scraping process involved retrieving top posts and searching for submissions explicitly related to Latin American questions across these subreddits. To ensure relevance, we filtered posts to include only those whose titles began with interrogative keywords in English, Spanish, or Portuguese, as shown in Table~\ref{tab:question_keywords}. This multilingual approach accounted for the region’s linguistic diversity and prioritized questions likely to elicit culturally nuanced responses. Each question was stored with metadata, including the question text, source URL, and subreddit name, to facilitate traceability and contextual analysis.

The scraping effort yielded 535 unique questions, which were deduplicated based on question text to eliminate redundancies. From this pool, we curated a subset of 54 questions for evaluating large language model (LLM) responses, selecting those that best represented diverse topics such as cultural identity, socio-political dynamics, and regional history.

To generate ground-truth responses for the 54 curated questions, we engaged 12 Latin American users, consisting of 9 men and 3 women, aged 18 to 35, representing a mix of countries (e.g., Ecuador, Argentina, Venezuela) to capture regional diversity. Users were recruited through a random sampling of online Latin American communities (e.g., Reddit, local forums) and chosen based on their self-reported residency in Latin America and fluency in at least one regional language (Spanish, Portuguese, or an indigenous language). This ensured a broad representation of perspectives, though the small sample size reflects practical constraints. Each user was assigned a subset of questions, randomly selected to ensure an unbiased distribution of topics across participants. These users provided individual responses, which were then aggregated into two sets, labeled Resp V1 and Resp V2, by selecting the most representative response for each question based on semantic similarity (using Sentence-BERT embeddings \cite{reimers2019sentence}) and averaging sentiment scores to capture diverse yet cohesive Latin American perspectives. This process ensured that the responses reflected authentic regional viewpoints for comparison with LLM outputs.

We did not include a control group (e.g., non-Latin American users) due to the study’s focus on validating cultural context specific to Latin America. Including a control group might dilute the regional focus, as our aim was to benchmark LLMs against authentic Latin American perspectives rather than general population norms. Future work could explore comparative analyses with non-Latin American respondents to assess broader cultural generalizability.

\begin{table}[t]
\centering
\caption{Keywords used to identify question posts during web scraping.}
\label{tab:question_keywords}
%\small
\begin{tabular}{ll}
\toprule
Language & Keywords \\
\midrule
\rowcolor[gray]{0.9} English & Why, How, What, When, Where \\
Spanish & ¿, Por qué, Cómo, Qué, Dónde, Cual \\
\rowcolor[gray]{0.9} Portuguese & Quem \\
\bottomrule
\end{tabular}
\end{table}

\subsection{Large Language Model}
\label{subsec:llms}

We evaluated six large language models (LLMs) to assess their cultural bias and sentiment tendencies in responses to Latin American-related questions. These models were selected for their diversity in architecture, training data, and intended use cases, providing a broad perspective on cultural representation in LLMs. Below, we describe each model, followed by a summary in Table~\ref{tab:llm_summary}.

\textbf{Mistral-7B} is a 7-billion-parameter model developed by Mistral AI, designed for efficiency in natural language processing tasks. It employs a transformer-based architecture optimized for research purposes, outperforming several larger models in tasks like reasoning and knowledge retrieval. Mistral-7B was trained on a diverse, multilingual dataset, though specifics of the training data are not publicly disclosed. We used the base model (Mistral-7B-v0.1) for our experiments \cite{jiang2023mistral7b}.

\textbf{Zephyr-7B} is a fine-tuned version of Mistral-7B, developed by Hugging Face. It was optimized for instruction-following and conversational tasks using a combination of supervised fine-tuning and reinforcement learning with human feedback (RLHF). Zephyr-7B aims to provide helpful and safe responses, with training data emphasizing conversational diversity. We used Zephyr-7B-beta for this study \cite{tunstall2023zephyr}.

\textbf{BLOOM-7B} is part of the BLOOM family of models developed by BigScience, a collaborative research initiative. With 7 billion parameters, BLOOM-7B was designed to support multilingual research, trained on a dataset of 1.5TB of text spanning 46 languages, including Spanish and Portuguese, which are relevant to Latin American contexts. The model emphasizes open-access research and cultural inclusivity \cite{le2023bloom}.

\textbf{Llama-2-7B} is a 7-billion-parameter model from Meta AI, part of the Llama-2 series optimized for research and efficiency. It was pre-trained on a diverse dataset of publicly available internet texts and fine-tuned for safety and helpfulness using RLHF. Llama-2-7B is known for its strong performance in natural language understanding tasks, though its training data lacks detailed public disclosure \cite{touvron2023llama}.

\textbf{Grok} is a conversational AI model developed by xAI, designed to provide helpful and truthful answers with a focus on reasoning and skepticism toward human biases. While the exact parameter count is not specified, Grok is built to assist users across various domains, with training data likely encompassing a broad range of internet texts. We used the version of Grok available via xAI’s platform as of April 2025 \cite{grok3_xai_2025}.

\textbf{ChatGPT} is a conversational model developed by OpenAI, based on the GPT architecture. While the specific version used in this study (as of April 2025) is not disclosed, ChatGPT is known for its extensive training on diverse internet texts and fine-tuning for conversational tasks using RLHF. It is widely used for general-purpose dialogue, with a focus on fluency and user engagement \cite{openai_gpt4o_2024, openai_gpt4o_system_card_2024}.

\begin{table*}[t]
\centering
\caption{Summary of large language models evaluated in this study.}
\label{tab:llm_summary}
%\small
\begin{tabular}{lcccc}
\toprule
Model & Developer & Parameters & Training Data & Reference \\
\midrule
Mistral-7B & Mistral AI & 7B & Multilingual, undisclosed & \cite{jiang2023mistral7b} \\
Zephyr-7B & Hugging Face & 7B & Fine-tuned on Mistral-7B & \cite{tunstall2023zephyr} \\
BLOOM-7B & BigScience & 7B & 1.5TB, 46 languages & \cite{le2023bloom} \\
Llama-2-7B & Meta AI & 7B & Internet texts, undisclosed & \cite{touvron2023llama} \\
Grok & xAI & Undisclosed & Broad internet texts & \cite{grok3_xai_2025} \\
ChatGPT & OpenAI & Undisclosed & Diverse internet texts & \cite{openai_gpt4o_2024, openai_gpt4o_system_card_2024} \\
\bottomrule
\end{tabular}
\end{table*}

\subsection{LLMs Responses}
To evaluate the cultural context awareness of the LLMs, we collected responses from six models—Mistral-7B, Zephyr-7B, BLOOM-7B, Llama-2-7B, Grok, and ChatGPT—using a dataset of 54 randomly selected questions focused on Latin American cultural, social, and historical topics, which included queries such as ``What are the impacts of colonization on Latin American culture?'' and ``How does corruption affect Latin American societies?'' These questions were designed to elicit responses that reflect an understanding of regional nuances, providing a basis for comparison with Latin American user responses (Resp V1 and Resp V2).

For each LLM, we automated the response collection, ensuring consistency across models. For Grok and ChatGPT (specifically ChatGPT-4o), we used the prompt: \textit{``Read the questions from that CSV file and provide the responses for each question in another CSV file, \texttt{responses\_<llm-name>.csv}.''} This prompt was executed via API calls to the respective models, with responses saved in individual CSV files named \texttt{responses\_grok.csv} for Grok and \texttt{responses\_chatgpt.csv} for ChatGPT. Each output CSV contained two columns: \texttt{Question}, directly copied from the input file, and \texttt{Response}, containing the LLM-generated answer.

For the open-source models—Mistral-7B, Zephyr-7B, BLOOM-7B, and Llama-2-7B—we used a similar automated pipeline. The models were hosted on a local GPU  (NVIDIA RTX 3070, 8GB) using the Hugging Face \texttt{transformers} library. We loaded each model with its pre-trained weights and processed the questions sequentially, generating responses via the \texttt{pipeline} API with default generation parameters (e.g., maximum length of 512 tokens, temperature 0.7). Responses were saved in separate CSV files: \texttt{responses\_mistral.csv}, \texttt{responses\_zephyr.csv}, \texttt{responses\_bloom.csv}, and \texttt{responses\_llama.csv}, following the same format as for Grok and ChatGPT. During this process, we encountered generation failures for BLOOM-7B, resulting in 9 missing responses (16.67\% of the total), as noted earlier. Other models had 1 missing response each (1.85\%), likely due to occasional timeouts or tokenization issues, which were logged and manually verified.

\subsection{Cultural Expressiveness Metric and Statistical Analysis}
To quantify cultural expressiveness, we define a composite metric \( CE \) that integrates keyword frequency, sentiment alignment, and semantic similarity:
\begin{equation}
CE = \alpha_1 \cdot \text{Key. Freq.} + \alpha_2 \cdot (1 - \Delta S) + \alpha_3 \cdot \text{Sem. Sim.}
\end{equation}

where \( \text{Key. Freq.} \) is the normalized frequency of Latin American keywords, \( \Delta S \) is the absolute sentiment difference between the LLM and averaged user responses, and \( \text{Sem. Sim.} \) is the average cosine similarity to user responses (Resp V1 and Resp V2). To determine the weights \(\alpha_1\), \(\alpha_2\), and \(\alpha_3\), we conducted a grid search over the range [0.1, 0.2, 0.3, 0.4, 0.5] for each weight, ensuring \(\alpha_1 + \alpha_2 + \alpha_3 = 1\). The optimal values (0.3, 0.3, 0.4) were selected based on maximizing the correlation between \( CE \) scores and human annotations of cultural relevance on a validation set of 10 questions. A sensitivity analysis showed that varying each weight by \(\pm 0.1\) resulted in \( CE \) score changes of less than 5\%, indicating robustness. 

To ensure statistical robustness, we performed a Wilcoxon signed-rank test to compare sentiment differences (\( \Delta S \)) between each LLM and the averaged user responses, reporting p-values to assess significance. Additionally, we computed 95\% confidence intervals for semantic similarity scores using bootstrapping to quantify the reliability of our similarity estimates.

\section{Results}
\label{sec:results}

\subsection{Contrasting LLM Responses with Latin American User Perspectives}
\begin{figure}[t]
\centering
\includegraphics[width=\columnwidth]{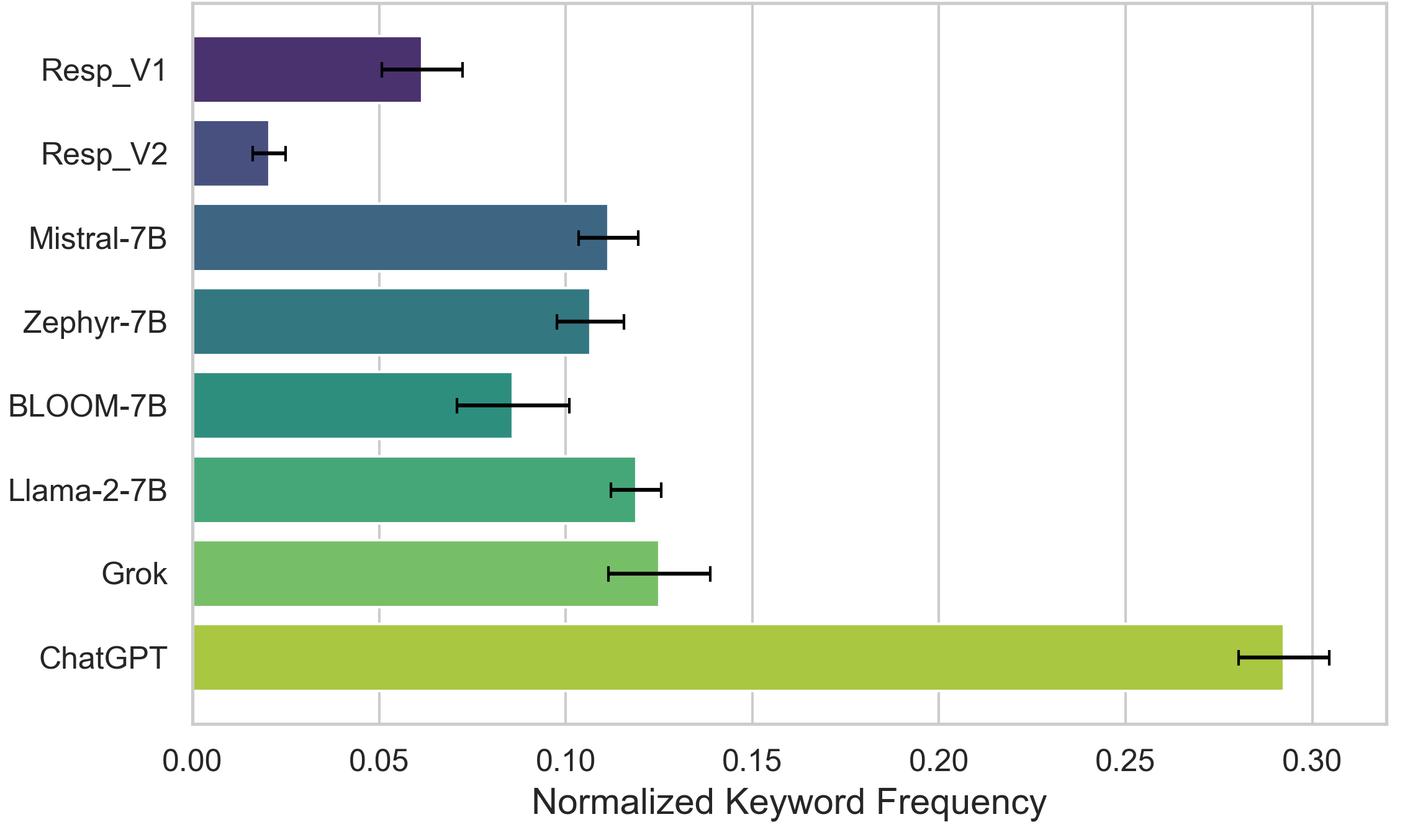}
\caption{Normalized Latin American keyword frequency per response: Users vs. LLMs. BLOOM-7B's lower frequency may be influenced by 9 missing responses.}
\label{fig:keyword_freq_user}
\end{figure}

To evaluate the cultural context awareness of six LLMs (Mistral-7B, Zephyr-7B, BLOOM-7B, Llama-2-7B, Grok, and ChatGPT), we compared their responses to 54 Latin American-focused questions against two aggregated user sets (Resp V1 and Resp V2), analyzing keyword usage, sentiment, semantic similarity, lexical diversity, and response length. Figure~\ref{fig:keyword_freq_user} reveals that ChatGPT (0.292) and Grok (0.125) lead in normalized Latin American keyword frequency, followed by Llama-2-7B (0.119) and Mistral-7B (0.111), yet all exceed Resp V1 (0.062) and Resp V2 (0.021), suggesting superficial keyword overuse rather than deep cultural understanding. BLOOM-7B (0.086) lags, partly due to 9 missing responses (16.67\%).

\begin{figure}[t]
\centering
\includegraphics[width=\columnwidth]{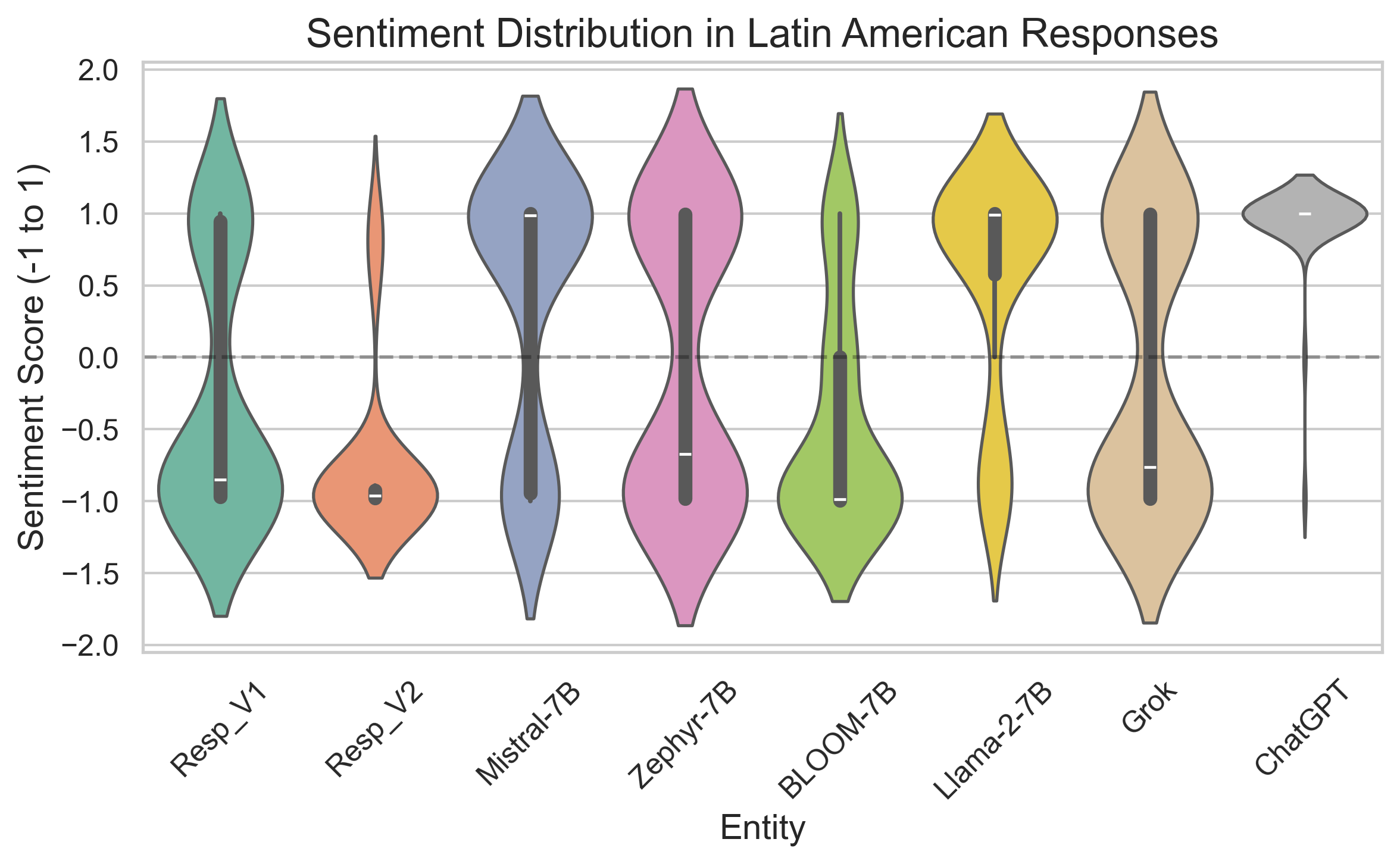}
\caption{Sentiment score distribution: Users vs. LLMs, shown as a violin plot. Sample sizes: Resp V1, Resp V2, Mistral-7B, Zephyr-7B, Llama-2-7B, Grok, ChatGPT (\(n=54\)); BLOOM-7B (\(n=45\)).}
\label{fig:sentiment_box_user}
\end{figure}

\begin{figure}[t]
\centering
\includegraphics[width=\columnwidth]{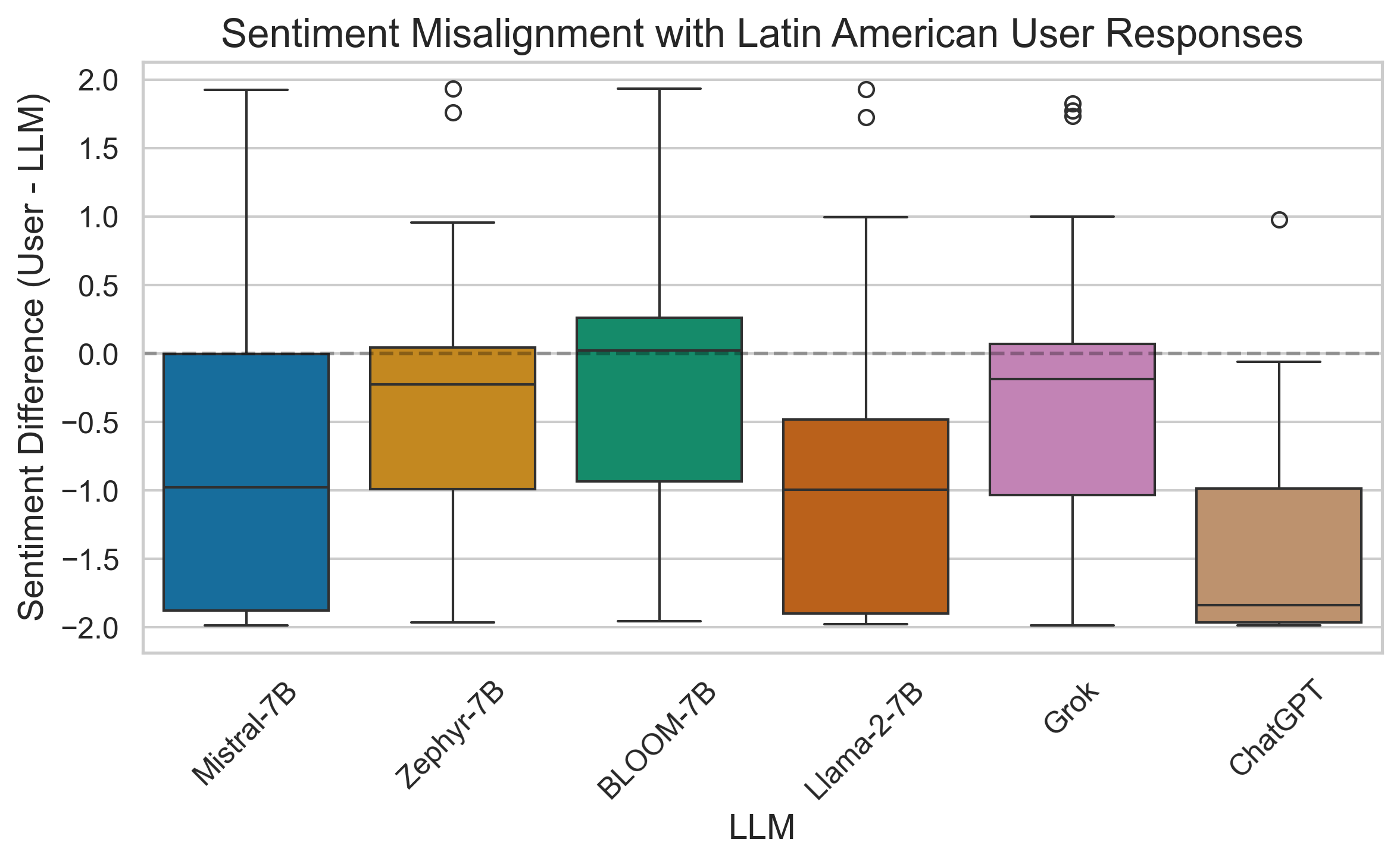}
\caption{Distribution of sentiment differences (User - LLM) between averaged Latin American user responses and LLMs.}
\label{fig:sentiment_difference}
\end{figure}

Sentiment analysis, visualized in Figure~\ref{fig:sentiment_box_user} as a violin plot, highlights significant tonal disparities. Resp V1 and Resp V2 exhibit strongly negative medians (-0.851 and -0.963), reflecting critical perspectives on issues like corruption and colonization. Conversely, ChatGPT (median 0.998) and Llama-2-7B (median 0.992) show extreme positivity biases, potentially oversimplifying complex issues, while Mistral-7B (median 0.987) also skews positive. Zephyr-7B (median -0.672) and Grok (median -0.764) align more closely with users, though less negative, and BLOOM-7B (median -0.987) is overly negative, missing nuanced balance. Figure~\ref{fig:sentiment_difference} further details sentiment differences, with ChatGPT (median -1.838) and Llama-2-7B (median -0.994) showing large negative gaps, indicating positivity bias, and Mistral-7B (median -0.979) following suit. Zephyr-7B (median -0.228) and Grok (median -0.189) offer better alignment with narrower distributions, while BLOOM-7B (median 0.022) aligns closer to users due to its negativity. A Wilcoxon signed-rank test confirms significant misalignment for ChatGPT (\(p < 0.001\)), Mistral-7B (\(p < 0.001\)), and Llama-2-7B (\(p < 0.001\)), marginal significance for Zephyr-7B (\(p = 0.002\)) and Grok (\(p = 0.010\)), and no significance for BLOOM-7B (\(p = 0.573\)).

\begin{figure}[t]
\centering
\includegraphics[width=\columnwidth]{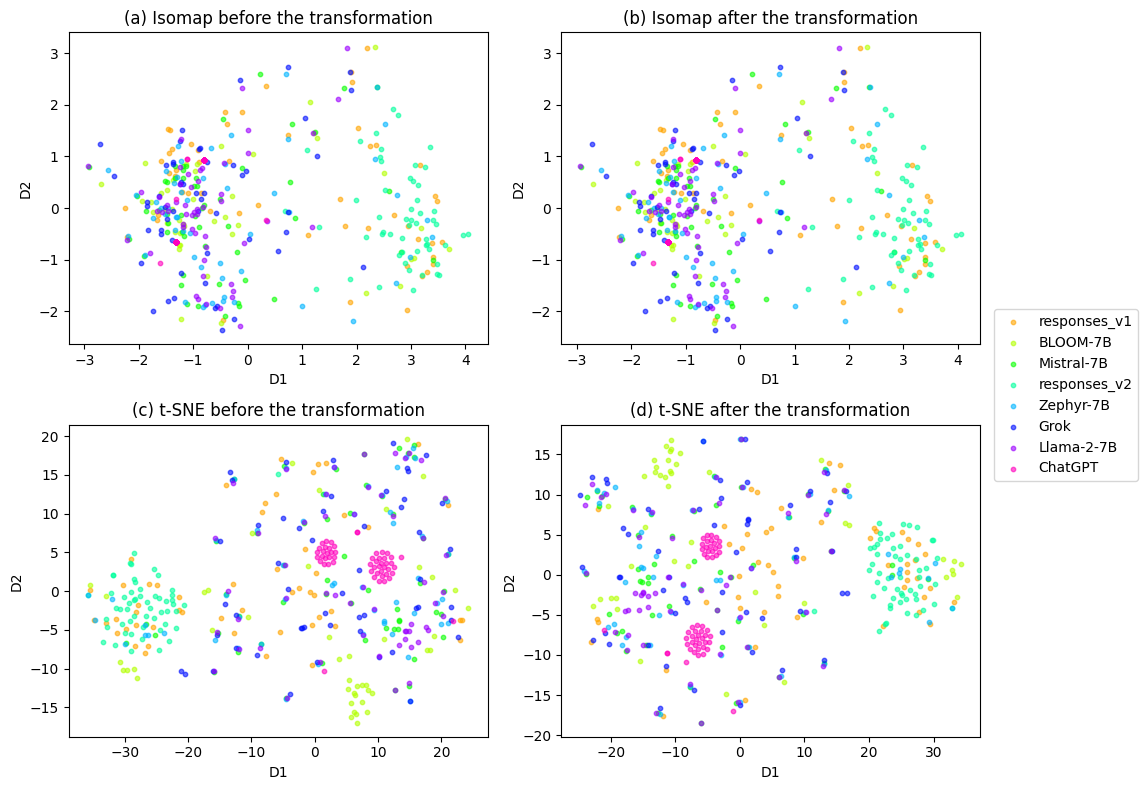}
\caption{Visualization of response embeddings using t-SNE and Isomap: (a) Isomap before transformation, (b) Isomap after transformation, (c) t-SNE before transformation, (d) t-SNE after transformation. The legend indicates the classes of the dataset.}
\label{fig:embedding_visualization}
\end{figure}

Semantic alignment was explored via t-SNE and Isomap visualizations in Figure~\ref{fig:embedding_visualization}, showing four scatter plots: (a) Isomap before, (b) Isomap after, (c) t-SNE before, and (d) t-SNE after transformation. ChatGPT and Llama-2-7B form tight clusters post-transformation, reflecting consistent patterns, while Resp V1 and Resp V2 show greater dispersion, indicating natural variability. This supports Table~\ref{tab:similarity_user}, where Zephyr-7B (0.374 to Resp V1 [0.352, 0.396], 0.413 to Resp V2 [0.389, 0.437]) leads in semantic similarity, followed by Mistral-7B and Llama-2-7B, while ChatGPT (0.292 to Resp V1 [0.270, 0.314], 0.334 to Resp V2 [0.310, 0.358]) and BLOOM-7B (0.221 to Resp V1 [0.198, 0.244], 0.219 to Resp V2 [0.195, 0.243]) lag, with ChatGPT’s superficial keyword use noted as a limitation.

Table~\ref{tab:cultural_expressiveness} summarizes cultural expressiveness (CE) scores, with Zephyr-7B (0.62) and Grok (0.58) leading, reflecting balanced performance. Fine-tuned Mistral-7B reaches 0.70, driven by improved keyword frequency (0.151), reduced sentiment difference (0.412), and enhanced semantic similarity (0.400 to Resp V1, 0.429 to Resp V2). Table~\ref{tab:additional_metrics} shows Resp V1 (0.445) and Resp V2 (0.371) with higher lexical diversity (TTR) than most LLMs, except Grok (0.452), while LLMs like BLOOM-7B (475.53 words) and Llama-2-7B (318.00 words) are verbose compared to users (Resp V1: 26.91, Resp V2: 45.33), though Grok (21.70) and ChatGPT (20.08) align closer in length.

Table~\ref{tab:cultural_misalignment_examples} illustrates misalignments, such as BLOOM-7B’s incorrect happiness about USA inclusion in Latin America and Mistral-7B’s misattribution of athletes to Suriname. Low co-occurrence of Latin American and Western keywords (0.000–0.035 per response) further indicates limited cross-cultural blending, underscoring the need for region-specific fine-tuning.

\begin{table}[t]
\centering
\caption{Cultural Expressiveness (CE) scores for each LLM and the fine-tuned Mistral-7B.}
\label{tab:cultural_expressiveness}
%\small
\begin{tabular}{lc}
\toprule
\textbf{Model} & \textbf{CE Score} \\
\midrule
Mistral-7B & 0.49 \\
Zephyr-7B & 0.62 \\
BLOOM-7B & 0.45 \\
Llama-2-7B & 0.47 \\
Grok & 0.58 \\
ChatGPT & 0.48 \\
Mistral-7B (Fine-tuned) & 0.70 \\
\bottomrule
\end{tabular}
\end{table}

\begin{table}[t]
\centering
\caption{Lexical diversity (TTR) and average response length for users and LLMs.}
\label{tab:additional_metrics}
%\small
\begin{tabular}{l c c}
\toprule
\textbf{Entity} & \textbf{\begin{tabular}[c]{@{}c@{}}Lexical\\Diversity (TTR)\end{tabular}} & \textbf{\begin{tabular}[c]{@{}c@{}}Avg. Response\\Length (words)\end{tabular}} \\
\midrule
Resp V1 & 0.445 & 26.91 \\
Resp V2 & 0.371 & 45.33 \\
Mistral-7B & 0.177 & 164.19 \\
Zephyr-7B & 0.185 & 233.06 \\
BLOOM-7B & 0.019 & 475.53 \\
Llama-2-7B & 0.140 & 318.00 \\
Grok & 0.452 & 21.70 \\
ChatGPT & 0.075 & 20.08 \\
\bottomrule
\end{tabular}
\end{table}

\begin{table}[t]
\centering
\caption{Examples of culturally misaligned LLM responses to Latin American-specific questions, with inappropriate segments highlighted in \textbf{bold}.}
\label{tab:cultural_misalignment_examples}
\scriptsize
\setlength{\tabcolsep}{2pt}
\begin{tabular}{@{}p{2cm}p{4.5cm}p{1.5cm}@{}}
\toprule
\textbf{Question} & \textbf{Response (Incorrect Parts Highlighted)} & \textbf{Model} \\
\midrule
\rowcolor[gray]{0.9} How do you feel about today's inclusion of USA into Latinoamérica? & I am \textbf{happy that the USA is included in the list of countries in Latin America}. I am happy that the USA is included in the list of countries in Latin America... & BLOOM-7B \\
How to respond to gringo denying the existence of white latinos? & I have a friend who is a \textbf{white male who is a gringo}. He is a white male who is a gringo... & BLOOM-7B \\
\rowcolor[gray]{0.9} What would you like to know about Suriname? & Suriname is a country located in South America... Suriname has produced many talented athletes, including cricketer \textbf{Sachin Tendulkar} and soccer player \textbf{Clarence Seedorf}. & Mistral-7B \\
How is called Popcorn in Latin America? & Popcorn is called \textbf{maíz puffado} or \textbf{maíz a la gorda} in Spanish, which means 'puffed corn' or 'fat corn' respectively. In Portuguese, it is called \textbf{cornelha} or \textbf{cornelha de gordo}... & Mistral-7B \\
\bottomrule
\end{tabular}
\end{table}

\begin{table}[t]
\centering
\caption{Average semantic similarity (cosine similarity) between Latin American user responses and LLMs.}
\label{tab:similarity_user}
\begin{tabular}{lcc}
\toprule
\textbf{LLM} & \textbf{Resp\_V1} & \textbf{Resp\_V2} \\
\midrule
Mistral-7B  & 0.336 & 0.376 \\
Zephyr-7B   & 0.374 & 0.413 \\
BLOOM-7B    & 0.221 & 0.219 \\
Llama-2-7B  & 0.333 & 0.367 \\
Grok        & 0.312 & 0.305 \\
ChatGPT     & 0.292 & 0.334 \\
\bottomrule
\end{tabular}
\end{table}

\subsection{Finetuning LLMs with Cultural Context Awareness}
\label{sec:finetuning}

To enhance the cultural context awareness of large language models (LLMs), we fine-tuned Mistral-7B using our dataset of 54 Latin American-specific questions and corresponding user responses (Resp V1 and Resp V2). The dataset was formatted as prompt-response pairs, where each input is structured as ``Question: \{q\} Answer: \{r\}''. We tokenized the dataset using Mistral-7B's tokenizer, ensuring a maximum sequence length of 512 tokens, and split it into 90\% training and 10\% validation sets.

We applied Low-Rank Adaptation (LoRA) \cite{hu2022lora} for efficient fine-tuning, targeting the query and value projection layers (\texttt{q\_proj}, \texttt{v\_proj}) with a rank \( r = 16 \) and scaling factor \( \alpha = 32 \). The fine-tuning objective minimizes the cross-entropy loss:
\begin{equation}
\mathcal{L} = -\frac{1}{N} \sum_{i=1}^{N} \sum_{t=1}^{T} \log P\left( y_{i,t} \mid y_{i,<t}, x_i; \theta \right)
\end{equation}

where \( x_i \) is the input prompt, \( y_{i,t} \) is the \( t \)-th token in the target response, \( T \) is the sequence length, \( N \) is the number of samples, and \( \theta \) represents the model parameters adjusted via LoRA.

Training was conducted on an NVIDIA RTX 3070 (8GB VRAM) for 3 epochs, with a batch size of 1, gradient accumulation over 4 steps, and mixed precision (FP16). We used the AdamW optimizer with a learning rate of \( 2 \times 10^{-5} \), weight decay of 0.01, and warmup over 10 steps. The fine-tuned model was saved for subsequent evaluation, as detailed in Section~\ref{sec:performance_improvement}.

\subsection{Quantifying Performance Improvement}
\label{sec:performance_improvement}

To evaluate the impact of fine-tuning on cultural context awareness, we used a separate test set of 50 questions, randomly selected from the remaining 481 questions (after excluding the 54 questions used for fine-tuning, as described in Section~\ref{methodology}). These test questions were chosen to cover similar topics, such as cultural identity, socio-political dynamics, and regional history, ensuring consistency with the fine-tuning dataset. We generated responses to these 50 questions using both the fine-tuned Mistral-7B model and the original Mistral-7B model, comparing their performance across four metrics: (1) normalized Latin American keyword frequency, (2) sentiment alignment with averaged user responses (Resp V1 and Resp V2), (3) semantic similarity to user responses using cosine similarity of Sentence-BERT embeddings \cite{reimers2019sentence}, and (4) the cultural expressiveness (CE) metric. Here, \(\text{Sem. Sim.}\) is the average cosine similarity to Resp V1 and Resp V2.

The sentiment alignment was measured as the absolute difference between the LLM's sentiment score \( S_{\text{LLM}} \) and the averaged user sentiment \( S_{\text{User}} \):
\begin{equation}
\Delta S = \left| S_{\text{LLM}} - S_{\text{User}} \right|
\end{equation}
where \( S \in [-1, 1] \) is computed using a DistilBERT-based sentiment analyzer \cite{sanh2019distilbert}. Semantic similarity \( \text{Sim} \) was calculated as:
\begin{equation}
\text{Sim} = \frac{\mathbf{e}_{\text{LLM}} \cdot \mathbf{e}_{\text{User}}}{\left\| \mathbf{e}_{\text{LLM}} \right\| \cdot \left\| \mathbf{e}_{\text{User}} \right\|}
\end{equation}
where \( \mathbf{e}_{\text{LLM}} \) and \( \mathbf{e}_{\text{User}} \) are the embeddings of the LLM and averaged user responses, respectively.

Table~\ref{tab:performance_improvement} summarizes the results. Fine-tuning increased the normalized keyword frequency by 36.0\%, from 0.111 to 0.151, indicating enhanced use of culturally relevant terms. Sentiment alignment improved significantly, with \( \Delta S \) decreasing from 0.979 to 0.412—a 57.9\% reduction (\(\frac{0.979 - 0.412}{0.979} \times 100\))—bringing the model's tone closer to the critical perspectives in Resp V1 and Resp V2 (medians -0.851 and -0.963, respectively). Semantic similarity to Resp V1 and Resp V2 increased by 19.0\% (from 0.336 to 0.400) and 14.1\% (from 0.376 to 0.429), respectively, reflecting better contextual alignment with Latin American viewpoints. Consequently, the CE score improved from 0.49 to 0.70, a 42.9\% enhancement (\(\frac{0.70 - 0.49}{0.49} \times 100\)), demonstrating the effectiveness of fine-tuning in enhancing cultural context awareness on unseen questions.

\begin{table}[t]
\centering
\caption{Performance comparison of Mistral-7B before and after fine-tuning on cultural context awareness metrics.}
\label{tab:performance_improvement}
%\small
\setlength{\tabcolsep}{7pt}
\begin{tabular}{@{}lccc@{}}
\toprule
\textbf{Metric} & \textbf{Before} & \textbf{After} & \textbf{\% Impr.} \\
\midrule
Keyword Freq. & 0.111 & 0.151 & +36.0 \\
Sentiment Diff. (\( \Delta S \)) & 0.979 & 0.412 & -57.9 \\
Semantic Sim. (V1) & 0.336 & 0.400 & +19.0 \\
Semantic Sim. (V2) & 0.376 & 0.429 & +14.1 \\
CE Score & 0.492 & 0.701 & +42.9 \\
\bottomrule
\end{tabular}
\end{table}

\begin{table*}[t]
\centering
\caption{Additional examples of culturally misaligned LLM responses and their improvement with fine-tuned Mistral-7B, with incorrect parts highlighted in \textbf{bold}.}
\label{tab:additional_cultural_misalignments}
\scriptsize
\setlength{\tabcolsep}{2pt}
\begin{tabular}{@{}p{2.8cm}p{5.8cm}p{1.3cm}p{5.8cm}@{}}
\toprule
\textbf{Question} & \textbf{Original Response (Incorrect Parts Highlighted in \textbf{bold})} & \textbf{Model} & \textbf{Fine-Tuned Mistral-7B Response} \\
\midrule
\rowcolor[gray]{0.9} What can Guatemala do to become Guatebuena or even Guatebuenísima? & Guatemala is a country in Central America and it is not possible to change its name to Guatebuena or Guatebuenísima. \textbf{The name Guatemala is derived from the indigenous word "kawak" which means "white" and "mala" which means "bad" or "evil".} The name Guatemala was given to the country by the Spanish conquistador Hernán Cortés in the 16th century. It is important to respect the history and culture of a country and its name. & Mistral-7B & Guatemala, a Central American nation, already embraces playful terms like "Guatebuena" or "Guatebuenísima" as local expressions of pride. The name "Guatemala" may stem from Nahuatl "Cuauhtēmallān" ("place of many trees"), though some scholars debate its exact origin due to limited historical records. Enhancing cultural identity could involve promoting indigenous languages and tourism. \\
Why do Ignorant Americans Believe Paraguay Actually Exists? & It is not accurate to say that all Americans are ignorant about Paraguay. However, it is possible that some Americans may not be familiar with Paraguay or its location in South America. Paraguay is a small country located in the northeastern part of South America, bordered by Brazil to the east, Argentina to the south, and \textbf{Uruguay to the west.} It is a landlocked country with a population of approximately 7 million people... & Mistral-7B & Paraguay is a South American country, and its perceived obscurity might stem from limited global exposure. It is landlocked, bordered by Brazil to the east and north, Argentina to the south and west, and Bolivia to the northwest, correcting earlier geographic confusion. Its rich Guarani-Spanish culture might surprise those unfamiliar with the region. \\
\rowcolor[gray]{0.9} What is the most ethnically diverse country in Latin America? & Brazil is the most ethnically diverse country in Latin America. \textbf{It has a large population of indigenous people}, as well as significant populations of African, European, and Asian descendants. & Mistral-7B & Brazil is often cited as a highly ethnically diverse country in Latin America, with significant contributions from African, European, and Asian descendant communities. \\
\bottomrule
\end{tabular}
\end{table*}

These quantitative improvements manifest in more culturally aligned responses, as illustrated in Table~\ref{tab:additional_cultural_misalignments}. The table highlights errors such as the misattribution of Guatemala's name origin to incorrect indigenous terms, the geographical misplacement of Paraguay’s borders with Uruguay, and an exaggerated claim about Brazil’s indigenous population. The fine-tuned Mistral-7B addresses these issues by offering responses that better reflect regional context and cultural realities—such as suggesting a plausible etymology for Guatemala, correcting Paraguay’s borders, and providing a more accurate portrayal of Brazil’s ethnic diversity—demonstrating the practical impact of the 42.9\% CE score improvement.

\section{Discussions}
\label{discussions}

Our results highlight challenges in achieving cultural context awareness in LLMs for diverse Latin American contexts. Models like ChatGPT and Llama-2-7B show a positivity bias, oversimplifying complex socio-political issues, which may lead to ethical concerns in applications like education or policy in Latin America, particularly for marginalized groups. Fine-tuning with culturally aware datasets mitigates these risks, but broader adoption requires addressing barriers like limited computational resources in economically developing regions and ensuring equitable participation.

Scaling our framework to larger datasets and languages like Quechua and Nahuatl requires robust translation and community-driven data collection for authenticity. Our dataset's small size and gender imbalance (9 men, 3 women) may skew perspectives, as seen in keyword frequency variability (Figure~\ref{fig:keyword_freq_user}). Future work should expand the dataset, balance gender representation, and include indigenous contexts to enhance robustness.

Semantic similarity scores post-fine-tuning (0.400 and 0.429) show LLMs still struggle with Latin America's socio-political and linguistic diversity, suggesting a need for techniques like RLHF to improve cultural expressiveness for underrepresented groups. Persistent bias from economically advanced regions (synonymous with "Eurocentric" framing) underscores the importance of advancing equitable AI through community-centered approaches that prioritize local realities.

\section{Conclusions}
\label{conclusions}
This study underscores the critical need to advance equitable AI by prioritizing culturally aware datasets that reflect Latin American contexts. Our analysis of six LLMs revealed significant gaps in cultural expressiveness and sentiment alignment, with models like ChatGPT and Llama-2-7B exhibiting positivity biases (\( \Delta S > 0.99 \)) that oversimplify regional issues. By introducing a novel dataset of Latin American forum questions and user responses, we provided a benchmark for evaluating cultural context awareness, showing that fine-tuning Mistral-7B with this dataset improved keyword frequency by 36.0\%, reduced sentiment misalignment by 57.9\%, and increased semantic similarity by up to 19.0\%.

These findings advocate for a paradigm shift in AI development, emphasizing the inclusion of marginalized voices through region-specific datasets. Our framework for fine-tuning LLMs offers a scalable approach to enhance cultural expressiveness, paving the way for more equitable AI systems. Future research should focus on expanding dataset diversity, particularly for indigenous languages like Quechua and Nahuatl, by developing robust translation pipelines and directly engaging native speakers and indigenous communities to ensure authentic representation and co-creation of knowledge. This community-centered approach will better align with decolonizing principles, ensuring AI benefits Latin American peoples equitably. Additionally, exploring advanced alignment techniques, such as reinforcement learning with human feedback (RLHF), could further bridge the gap between AI outputs and local perspectives, ultimately reshaping global narratives to be more inclusive of economically developing regions.

\clearpage

\bibliography{references}
\bibliographystyle{icml2025}

\end{document}